# The Method of Recursive Counting: Can One Go Further? *

M. Creutz[a], I. Horvath[a,b] and R. Mendris[c]

[a]Brookhaven National Laboratory, Upton, NY 11973, USA

[b] Physics Department, University of Rochester, Rochester, NY 14627, USA

[c] Department of Mathematics, Faculty of Electrical Engineering,
Slovak Technical University, 812 19 Bratislava, Slovakia

After a short review of the Method of Recursive Counting we introduce a general algebraic description of recursive lattice building. This provides a rigorous framework for discussion of method's limitations.

Recently there has been increased interest within the lattice community in high and low temperature series expansions. For discrete systems, very high orders are now available due to several developments including: 1) The Modified Shadow Method [1] is a computerized diagrammatic technique. It seems to be the most efficient way to calculate low temperature expansions in Potts-like models. 2) The Finite Lattice Method [2] relies on extracting the series expansion from the exact partition functions of small systems. 3) The Method of Recursive Counting (MRC) [3-5] is quite similar in spirit, but different in technical realization. Both methods are roughly equal in strength for Potts-like models and are probably the most powerful strategies for obtaining high temperature expansions for these systems.

In this presentation we pinpoint some of the essential ingredients in MRC. Building upon that, we introduce an algebraic formalism, which gives a rigorous means to discuss some of the limitations of MRC. We feel that our approach could eventually be useful in other contexts and might be interesting in its own right.

The heart of the MRC is Binder's proposal [6] for calculating the partition function of a finite system recursively. Consider an infinite 2-dimensional hypercubic grid without any variables on it. We build the system by adding finite straight lines (transverse layers) of variables

*Talk presented at the Lattice'93 conference in Dallas, Texas by Ivan Horvath. Preprint BNL-49783.

on top of each other. The last layer put into the bare grid is called the "exposed layer" and the ones below it are referred to as "covered layers".

Let $p(E, I)$ denote the number of states of our finite system, corresponding to given energy $E$ and a given configuration $I$ of the exposed layer. When adding a new layer, these numbers, usually called "counts", change in the obvious way

$$p(E, I) \longrightarrow p'(E, I) = \sum_{I'} p(E - \Delta(I, I'), I'). \quad (1)$$

Here $\Delta(I, I')$ is the change in the energy when configuration $I'$ is covered by the configuration $I$. Assuming that the energy of the system can only take discrete values and it is possible to keep the counts in the computer memory, we can add as many layers as we wish and in the end compute the partition function for the finite system we just built, namely

$$Z = \sum_E P(E) e^{-\beta E}, \quad (2)$$

where $P(E) = \sum_I p(E, I)$. Note that $e^{-\beta}$ is a low temperature variable and that the main advantage of this recursive procedure is that it saves one dimension in the problem.

With periodic boundary conditions in the transverse direction [3], adding new layers doesn't change the boundaries of the system - they are only at the top and the bottom. Therefore, if one subtracts the results for different volumes, finite size effects coming from excitations smaller than the length of the lattice cancel themselves out except for those that feel the periodicity in the transverse direction. Thus, such a subtraction represents a pleasantly simple procedure of extracting the low temperature expansion. If the system is substantially long, then the order of this expansion is completely specified by the "size" of the lattice in the transverse periodic subspace.

A powerful technical realization of the above strategy uses so called helical lattices [4]. The essential point lies in enumerating the variables as one puts them into the grid one after another, one at the time. Therefore, one can think of variables as lying on a chain. A helical lattice in two dimensions is specified by two positive integers $\{h_x, h_y\}$ representing the distance from a given variable on the chain to its $x, y$ neighbors. Using the convention that $h_y > h_x$, the exposed layer is formed by the last $h_y$ variables.

This formalism generates periodic boundary conditions in transverse direction automatically. To get from a given variable to its image, we must jump $n_x$ times in x-direction and $n_y$ times in y-direction and end up at the same place on the chain, i.e. $n_x h_x + n_y h_y = 0$. The vector $\vec{n} \equiv (n_x, n_y)$ characterizes the periodicity of the lattice. The order of the series expansion one can get from such a lattice is proportional to $L = |n_x| + |n_y|$. Clearly, we try to get the largest possible $L$ with the smallest number of variables in the exposed layer. It is easy to be optimal on the space of helical lattices: For fixed $h_y$, the best choice is $h_x = h_y - 1$ with $L = 2h_y - 1$. Note that this is twice as much as if we used a simple periodic lattice [5].

This is a very powerful result, but it makes one wonder: Are there even smarter lattices one could use for recursive counting in two dimensions? Turning to that question, we need to clarify the notion of the lattice, which is a bit confusing: We always mean the hypercubic grid and and yet talk about different "lattices". What the word lattice mainly reflects here is the relation of the set of variables to the grid, not the underlying geometric structure. More precisely, to specify a lattice means to specify a mapping from $Z \times Z$ to $Z$, where $Z \times Z$ represents the grid and $Z$ the set of variables. Note that since the integers form an ordered set, with a "lattice" in hand we not only know what the resulting system looks like, we also have a prescription to build it. In other words, we know in what order and where to put the variables on the grid. As we learned from Binder, the right choice of the building prescription can make a big difference here.

Inspired by helical lattices, we introduce the functions $x(n), y(n)$ representing the relative positions in $Z$ of $+x, +y$ neighbors of variable $n$ with respect to itself. Therefore, $p(n) \equiv n + x(n)$ is the $+x$ neighbor of variable $n$ and $q(n) \equiv n + y(n)$ its $+y$ neighbor. Clearly, not all the pairs $\{x(n), y(n)\}$ represent a lattice, and if such a description is supposed to be of any practical use, we should be able to tell when this is the case. In other words we have to define the set of lattices relying on the algebraic aspects only. This can be done.

Consider an elementary square on our grid and the variable $n$, residing in its lower left corner. From there to get to the diagonally opposite one, we can either go first up and then right, reaching the variable $p(q(n))$ or first right and then up, ending up at $q(p(n))$. To avoid contradiction on which variable actually resides there, commuting mappings $p(n), q(n)$ are required:

$$p(q(n)) = q(p(n)). \qquad (3)$$

While the unique existence of $+x, +y$ neighbors of a variable $n$ is guaranteed by the very existence of the functions $x(n), y(n)$, this is not necessarily the case for the the $-x, -y$ neighbors. This forces us to restrict ourselves to the invertible functions $p(n), q(n)$. In summary therefore, the pair $\{x(n), y(n)\}$ represents a lattice if and only if the corresponding $p(n), q(n)$ are one to one commuting mappings of some $B \subset Z$ on itself.

For example, a very broad class of algebraic lattices is described by functions: $y(n) = K$, $x(n + K) = x(n)$, where $K$ is an integer constant. Indeed, in this periodic case the commutation rule (3) is satisfied and we just restrict ourselves to injective $p(n), q(n)$.

For the purposes of the recursive procedure, we have to identify the "exposed layer". Quite generally, the variable belongs to the exposed layer, if some of their neighbours are not yet in place because the knowledge of such a variable is necessary to continue the recursion. In practice, we are restricted by computer memory and therefore to some finite number of variables in the exposed layer. In the algebraic language, the functions $x(n), y(n)$ are bounded by some positive constant $M$ i.e. $|x(n)|, |y(n)| \leq M, \forall n$. Indeed, if that is the case and $n$ is the last variable we put on the grid, then all of the variables up to $n - M$ have their neighbors certainly in place, leaving at most $M$ of them in the exposed layer.

Now going back to the question of optimal prescription for extracting the series expansion recursively, we can state it in our algebraic language: Consider the set of all algebraic lattices $\{\{x(n), y(n)\}\}_M$, for which $x(n), y(n)$ are bounded by $M$. It can be shown that for each lattice there is a unique vector $\vec{n} \equiv (n_x, n_y)$, representing the minimal periodicity of the lattice. The length $L$ of this vector in the Manhattan metric defines a functional on the above set. Find the lattice, for which the maximum of this functional is realized.

**Theorem:** *The maximum $L_{max} = 2M - 1$ is realized on the helical lattice $\{x(n) = M - 1, y(n) = M\}$.*

*Proof:* We give only a short sketch of the proof, relying on the lemma, which we won't prove. Having an arbitrary lattice $\{x(n), y(n)\}$, we can divide the set of variables $B$ into equivalence classes, each class consisting of variables, lying on the same line in x-direction.

**Lemma:** *If $x(n)$ is bounded by $M$, there are at most $M$ such equivalence classes. The maximum is achieved if and only if $x(n) = const = M$.*

The first part of the lemma tells us that for each lattice in our set, there are at most $M$ different lines of $x-$neighboring variables. Therefore, the lines repeat themselves on the grid and the repetition occurs after at most $M$ steps in the $y-$direction. In other words, to get from a given variable to its image, one has to take at most $M$ steps in $y-$direction. Now, exchanging the role of $x$ and $y$, we can say the same thing about hopping in $x-$direction and set the bound: $L_{max} \leq 2M$. The proof can now easily be completed, using the second part of the lemma. Putting $x(n) = y(n) = M$ we obtain the trivial case with periodicity $\vec{n} = (-1, 1)$ and the above bound can not be achieved. Therefore, the maximum is $L_{max} = 2M - 1$, because this value can be realized on the helical lattice $\{x(n) = M - 1, y(n) = M\}$. Q.E.D.

What we found is kind of a NO-GO theorem, because it asserts that what was already used is the best. There is a little hole though. Our algebraic description captures all the lattices for which the neighborhood of a given variable looks the same anywhere in the grid, including the orientation. In other words if a variable has an image somewhere, then there are also its image neighbors (as it should if we consider the local actions), with the same relative position. There do exist however also lattices for which the orientation of neighbors is not preserved. On the other hand, there are strong indications that the space of these lattices is quite small and that they are much less effective than the algebraic lattices.

The extension of our formalism to higher dimensions is straightforward. In those cases however, the MRC relies heavily on the cancelation tricks between different lattices [5] and the power of single lattice is not the only relevance there. On the other hand our algebraic approach enlarges the set of lattices with the potential of cancelling the errors and could therefore improve the efficiency of the method. This possibility has not been seriously investigated.

Lastly, although the result of this algebraic exercise is a NO-GO theorem, the fact that we were able to put this geometric problem in purely algebraic terms is interesting in itself. Our formalism might be useful in "algorithmization" of various tasks concerning finite lattices.